\documentclass[prb,twocolumn]{revtex4-2} 

\usepackage{float}
\usepackage{units}
\usepackage{graphicx}
\usepackage{hyperref}

\newcommand{\be}{\begin{equation}}
\newcommand{\ee}{\end{equation}}
\newcommand{\bea}{\begin{eqnarray}}
\newcommand{\eea}{\end{eqnarray}}

\begin{document}

\title{Don't throw that video away!\\ Reference Frames can fix Video Analysis with a Moving Camera}

\author{Nathan T. Moore}
\email{nmoore@winona.edu} % optional
\affiliation{Department of Physics, Winona State University, Winona, MN 55987}

\date{\today}

\begin{abstract}
One common source of error in video analysis is camera movement.  The paper describes a simple frame of reference correction that students can employ to salvage otherwise corrupted video analysis data.  Two examples are provided. 
\end{abstract}

\maketitle % title page is now complete

\section{Introduction} 
Video analysis is a convenient and fun way to collect kinematic position-time information for a motion that may not be otherwise accessible in the introductory lab.  The general procedure is to track the successive motion of an object via it's x and y pixel position over multiple frames of a video.  

One curriculum that uses video analysis as a learning support to construct and test ideas is Eugenia Etkina's ISLE approach to learning physics \cite{ISLE_overview}. 
For example, in the 2-D Projectile motion unit, one video \cite{ISLE_ball_video_source}, involves Dr. Etkina tossing a ball vertically while rolling across the room on rollerblades. If you are reading the paper on a computer, you might watch the video now \url{http://islevideos.net/experiment.php?topicid=2&exptid=95}.   

\begin{figure}[h]
\centering
\includegraphics[width=\columnwidth]{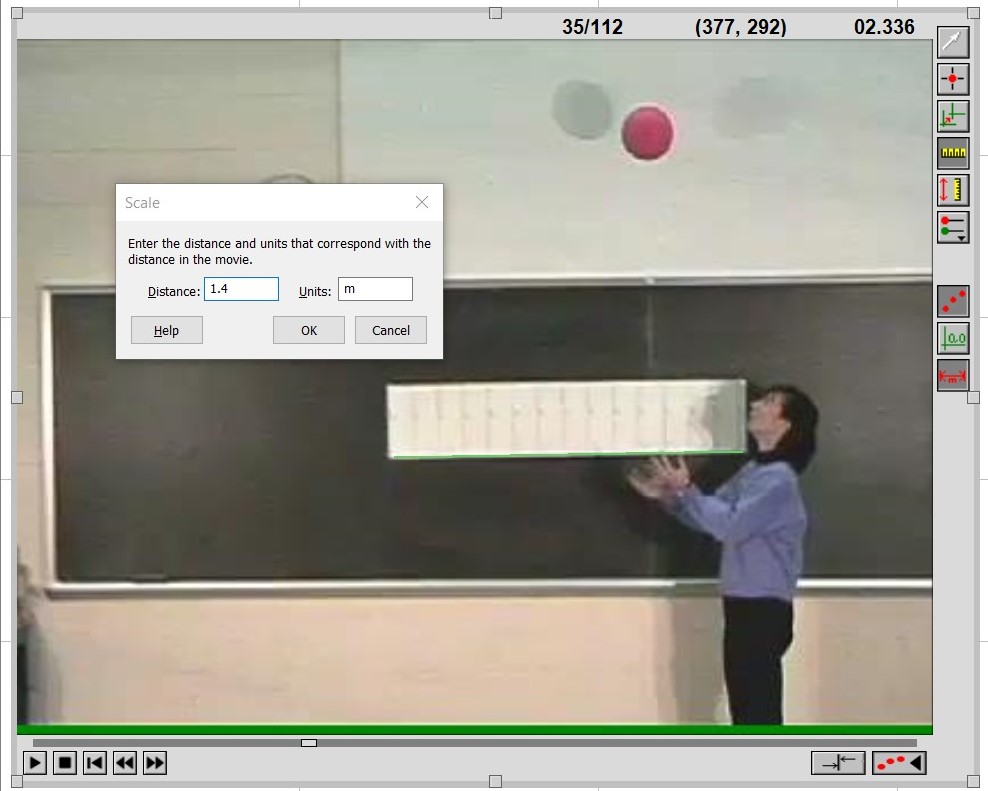}
\caption{ 
A screenshot from the ISLE-based video website (developed by E. Etkina and D.T.Brookes).
The video has been imported for analysis in Vernier's LoggerPro software.    
Horizontal  scale is given via the set of vertical lines drawn on the chalkboard with assumed spacing of $10$cm. 
The (green) calibration stick is assumed to be 1.4m in length via lines drawn on the chalkboard.  The video can be downloaded from \cite{ISLE_ball_video_source} .
}
\label{Etkina-calibration}
\end{figure}

\begin{figure}[h]
\centering
\includegraphics[width=\columnwidth]{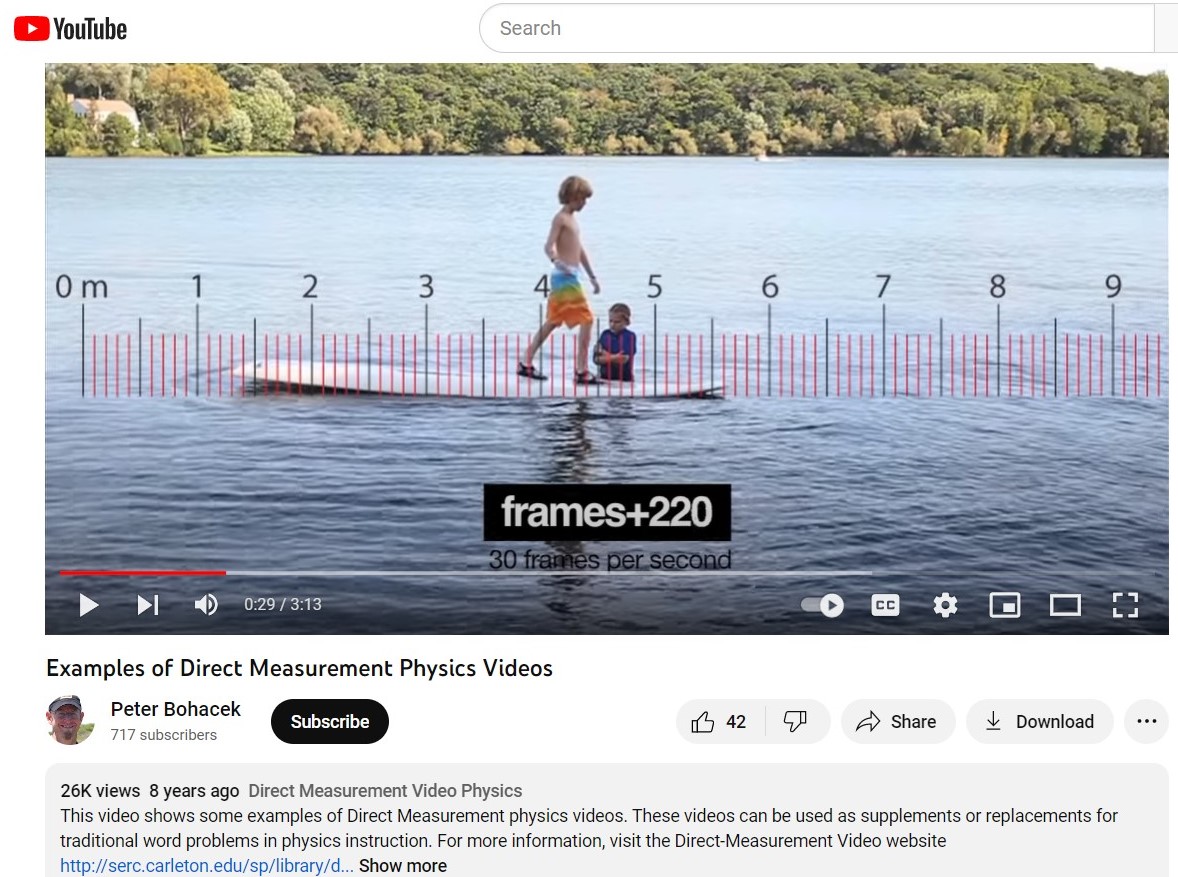}
\caption{A screenshot from Peter Bohacek's Direct Measurement Videos, \cite{Bohacek_youtube_intro}. In this case, position information is given by a drawn overlay added to the video via post-processing.
}
\label{Bohacek-1}
\end{figure}

Another set of teaching videos is Peter Bohacek's Direct Measurement Physics Videos, \cite{Bohacek_overview} .
These videos usually show physics in a ``real" setting (outside the classroom) and are again powerful and engaging tools.  

In both of these examples, students are presented with videos that can be analyzed frame by frame at constant time intervals, typically $60~frames/second$.  Horizontal and vertical positions are available by either marks on a chalkboard (Etkina) or a computer drawn overlay (Bohacek).  

There are also a number of video analysis software programs that allow students to analyze any old video they find or record with their cellphones.  ``Tracker'', \cite{Tracker}, is a free tool that works on a variety of platforms. Vernier's Logger Pro, \cite{LoggerPro}, often used for data acquisition, also contains video analysis software.  

% probably not needed
%\begin{figure}[h]
%\centering
%\includegraphics[width=\columnwidth]{Etkina-calibration.jpg}
%\caption{
%A screenshot from LoggerPro, showing a (green) $1.4$m calibration stick via lines on the chalkboard.
%}
%\label{Etkina-calibration}
%\end{figure}

In both of these packages, students repeatedly click on an object of interest and the software then collects time and 2D position data  in spreadsheet form.  However, the data collected is in units of an x,y pixel pair.  Conversion from pixel to physical dimension is a student task, and the quality of student results can depend on the ``calibration stick''\cite{calibration_stick} they employ.

\section{Fixing moving camera motion} 
\begin{figure}[h]
\centering
\includegraphics[width=\columnwidth]{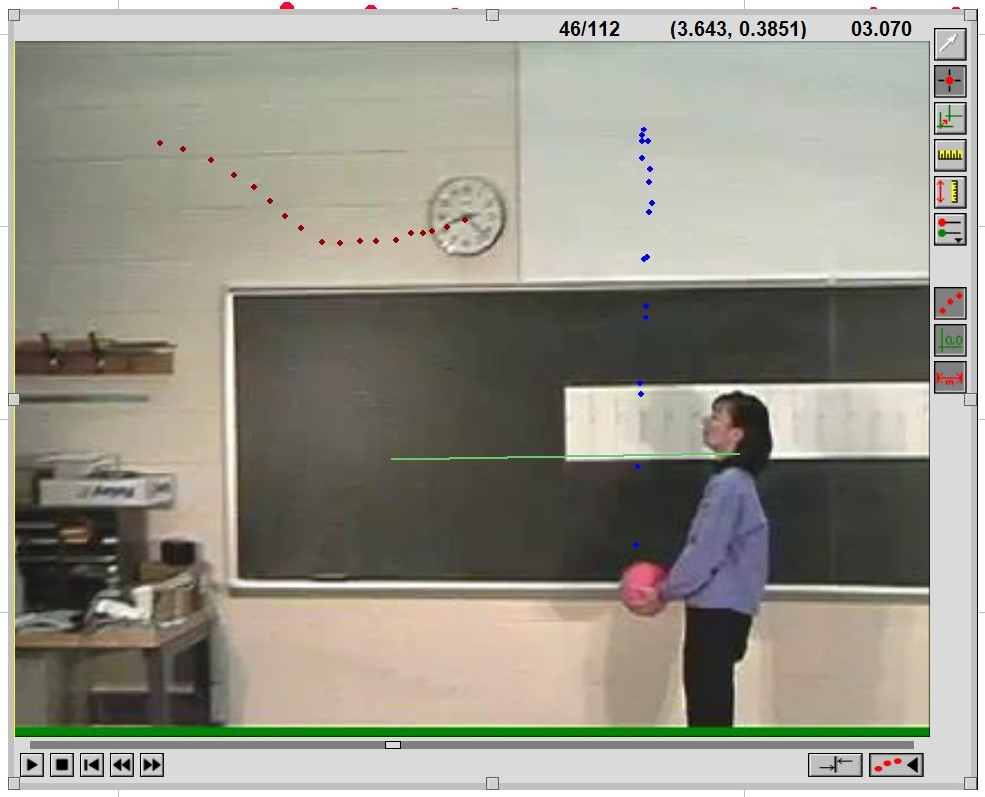}
\caption{
A screenshot from LoggerPro, showing both the track of the pink ball (blue dots) and the track made by the clock (red dots).  Note, the clock is attached to a concrete block wall, so the red dots also show the motion of the camera.  
While the ball is certainly moving across the classroom horizontally, the moving camera does not show this motion in the track of blue dots.
The position of the ball \textit{with respect to the classroom} can be generated via vector subtraction of the ball-camera and clock-camera reference frames.
}
\label{Etkina-dots-2}
\end{figure}

Figures \ref{Etkina-calibration} and \ref{Etkina-dots-2} show the analysis process for the video in which Dr. Etkina throws a ball vertically while rollerblading across the classroom.  The video of this event was taken by a camera that followed Etkina, so via direct tracking of the ball, the horizontal component is lost.

However, if a student also tracks the motion of a seemingly immobile object, for example the center seam of the chalkboard, the center of the clock, a corner of the calibration sheets, etc, the student can recover the horizontal motion of the ball via vector subtraction.  Specifically, if you write  the ball's position relative to the classroom wall as, $X_{ball~wrt~wall}$, you can then express this position as a difference.
\bea
X_{ball~wrt~wall} &=& X_{ball~wrt~camera}-X_{clock~wrt~camera} \nonumber\\ 
Y_{ball~wrt~wall} &=& Y_{ball~wrt~camera}-Y_{clock~wrt~camera}\nonumber
\eea
This subtraction can be accomplished in LoggerPro as a ``calculated column'' or the data can be exported to a spreadsheet and the operation performed there.  

\begin{figure}[h]
\centering
\includegraphics[width=\columnwidth]{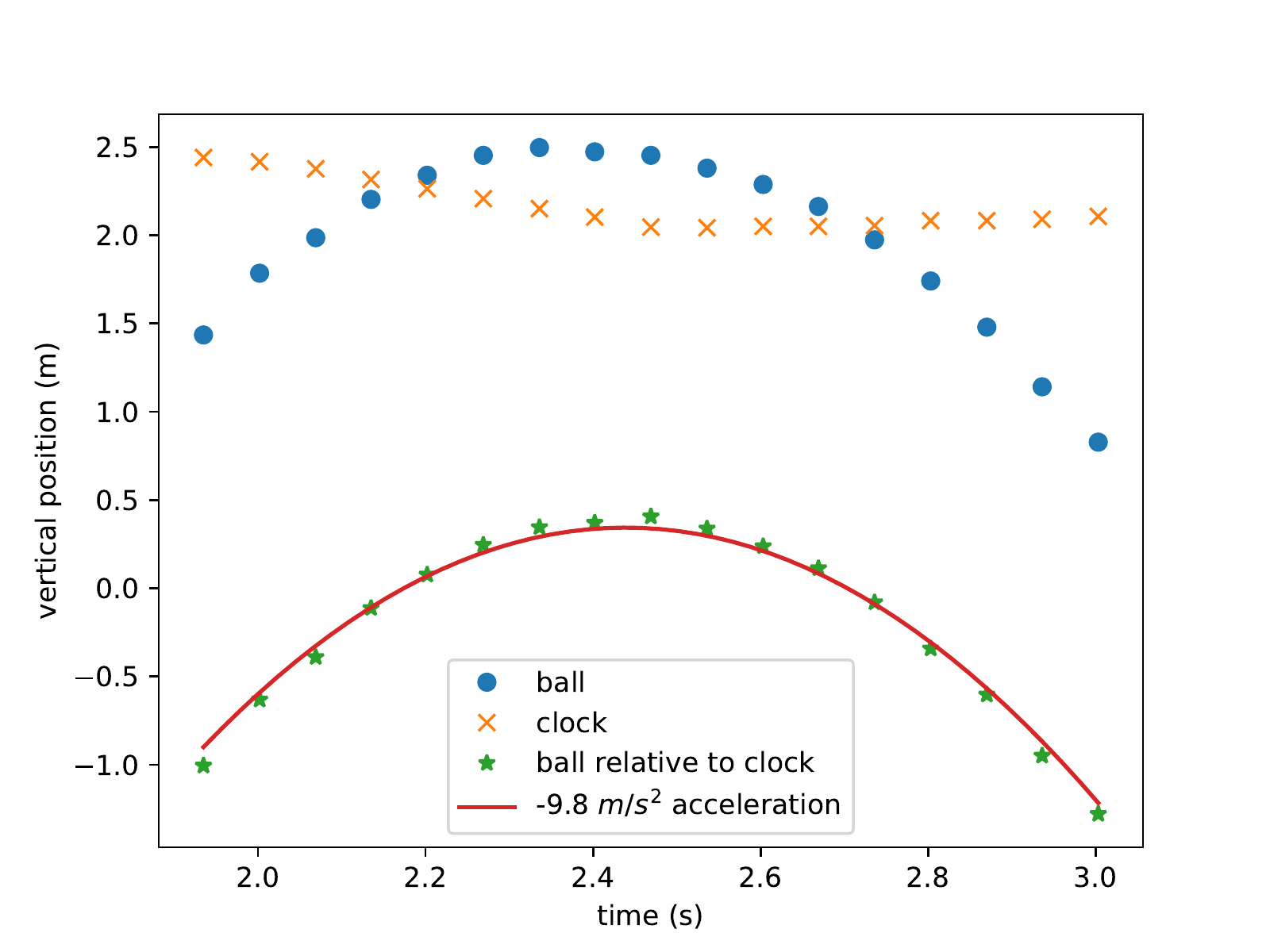}
\caption{
%A plot from LoggerPro showing the vertical position of the clock (red), the ball (blue), and the difference between the two (purple), the vertical position of the ball relative to the clock. Looking carefully, you can see that correcting for the relative motion of the camera eliminates the asymmetry in the (blue) direct measurement of altitude.  
A plot showing the vertical position of the clock, the ball, and the difference between the two, the vertical position of the ball relative to the clock. 
Looking carefully, you can see that correcting for the relative motion of the camera eliminates the asymmetry in the direct measurement of altitude.  
}
\label{Etkina-Y-T-plot}
\end{figure}

\begin{figure}[h]
\centering
\includegraphics[width=\columnwidth]{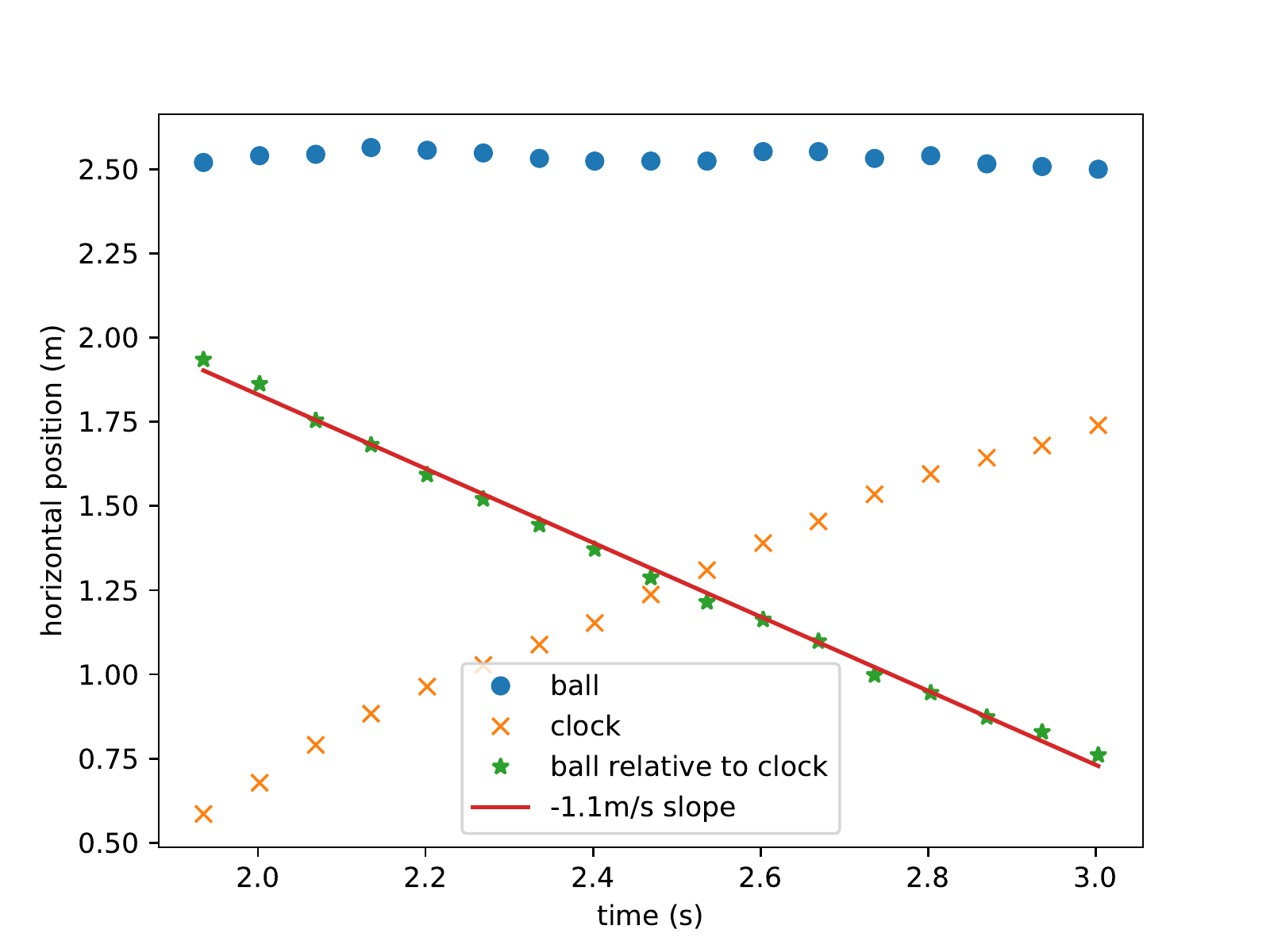}
\caption{
%A plot from LoggerPro showing the horizontal position of the clock (green), the ball (red), and the difference between the two (orange), ie the horizontal position of the ball relative to the clock.  This graph shows how motion hidden by a moving camera can be recovered if you subtract away the (camera) motion of a fixed reference.
A plot from LoggerPro showing the horizontal position of the clock, the ball, and the difference between the two, ie the horizontal position of the ball relative to the clock.  This graph shows how motion hidden by a moving camera can be recovered if you subtract away the (camera) motion of a fixed reference.
}
\label{Etkina-X-T-plot}
\end{figure}

Results from this change of reference frame are shown in figures \ref{Etkina-Y-T-plot} and \ref{Etkina-X-T-plot} . 
In the introductory curriculum, frames of reference sometimes seems dry or contrived, but in video analysis projects with a moving camera, thinking about the movement of a seemingly immobile reference frame can be a magic bullet.

% this figure probably isn't necessary
%\begin{figure}[h]
%\centering
%\includegraphics[width=\columnwidth]{Etkina-dots-1.jpg}
%\caption{
%A screenshot from LoggerPro, showing a track of blue dots from repeated clicking on the pink ball.  Position-time data is automatically recorded in spreadsheet form.
%}
%\label{Etkina-dots-1}
%\end{figure}

\clearpage

\section{Example: a bear falls out of a tree}
To further illustrate how useful this approach can be, consider this 2003 video of a bear being removed from a tree in Missoula Montana via a tranquilizer gun and a trampoline. \url{https://www.youtube.com/watch?v=9KiJnTGoPPI} \cite{bear_video_source}.  
There are quite a few different application topics available in the video, and some of my students are outraged that the bear was so made fun of by the Missoula fire, wildlife, and police departments.  

So, an ethical question: ``Assuming the bear had to be removed from the tree, was bouncing it off a trampoline a humane thing to do?''  With encouragement, the students can develop this question into a quantifiable measure, eg, ``What would the bear's speed be if there was no trampoline in place?''  There are many graphs online that show the fatality risk for pedestrians who are struck by cars at different speeds, for example \cite{AccidentRisk}, and it seems realistic to extrapolate this to a decrease in harm to the bear if its final velocity is reduced.

\begin{figure}[h]
\centering
\includegraphics[width=\columnwidth]{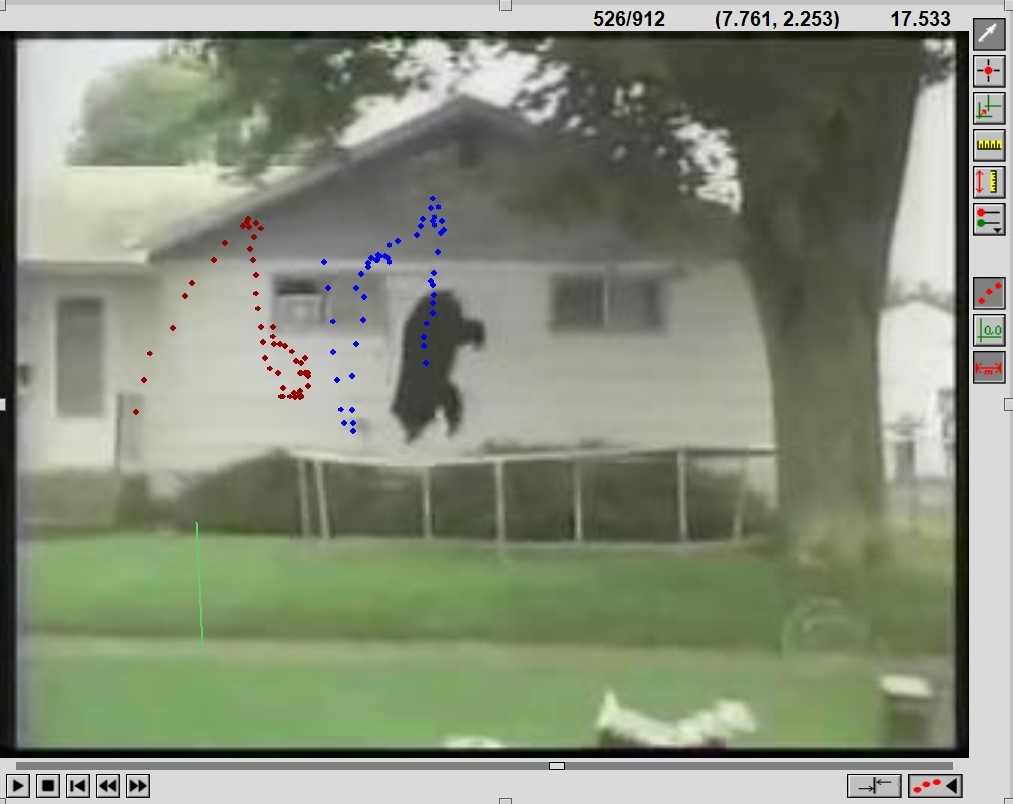}
\caption{
A screenshot from LoggerPro, showing the position-time track for both the bear and the lower left corner of the house's left transom window. The calibration used is the trampoline's leg height of $1$m.  The calibration stick has drifted off to the left because of the movement of the camera.  The calibration length is only an estimate and should probably be back fit to the appropriate gravitational acceleration value for the bear's free-fall.
}
\label{bear-dots}
\end{figure}

The bear video was shot by a professional videographer, Mark Hoyoak, and the focus of the camera follows the bear.  Ignoring the early parts of the bear's fall when the zoom level changes, the video provides a falling body, tracked by a moving camera.  If you consider one of the house's transom windows to be a static reference, camera movement can be removed.

\begin{figure}[h]
\centering
\includegraphics[width=\columnwidth]{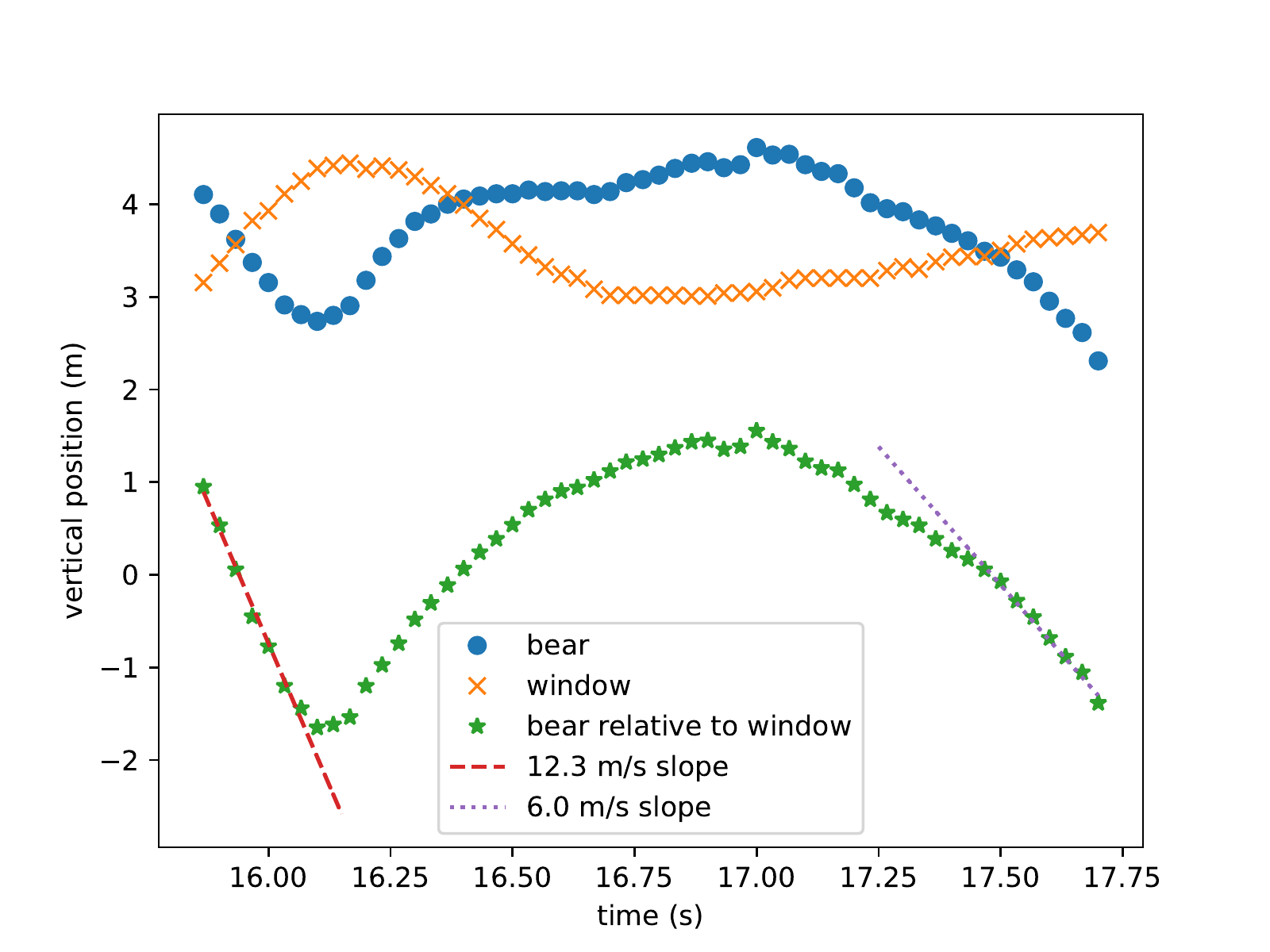}
\caption{
%A screenshot from LoggerPro, 
A plot showing the vertical position-time graph for both the bear, the lower left corner of the house's left transom window and the bear's (calculated)  motion relative to the window. 
}
\label{bear-speed}
\end{figure}

\begin{figure}[h]
\centering
\includegraphics[width=\columnwidth]{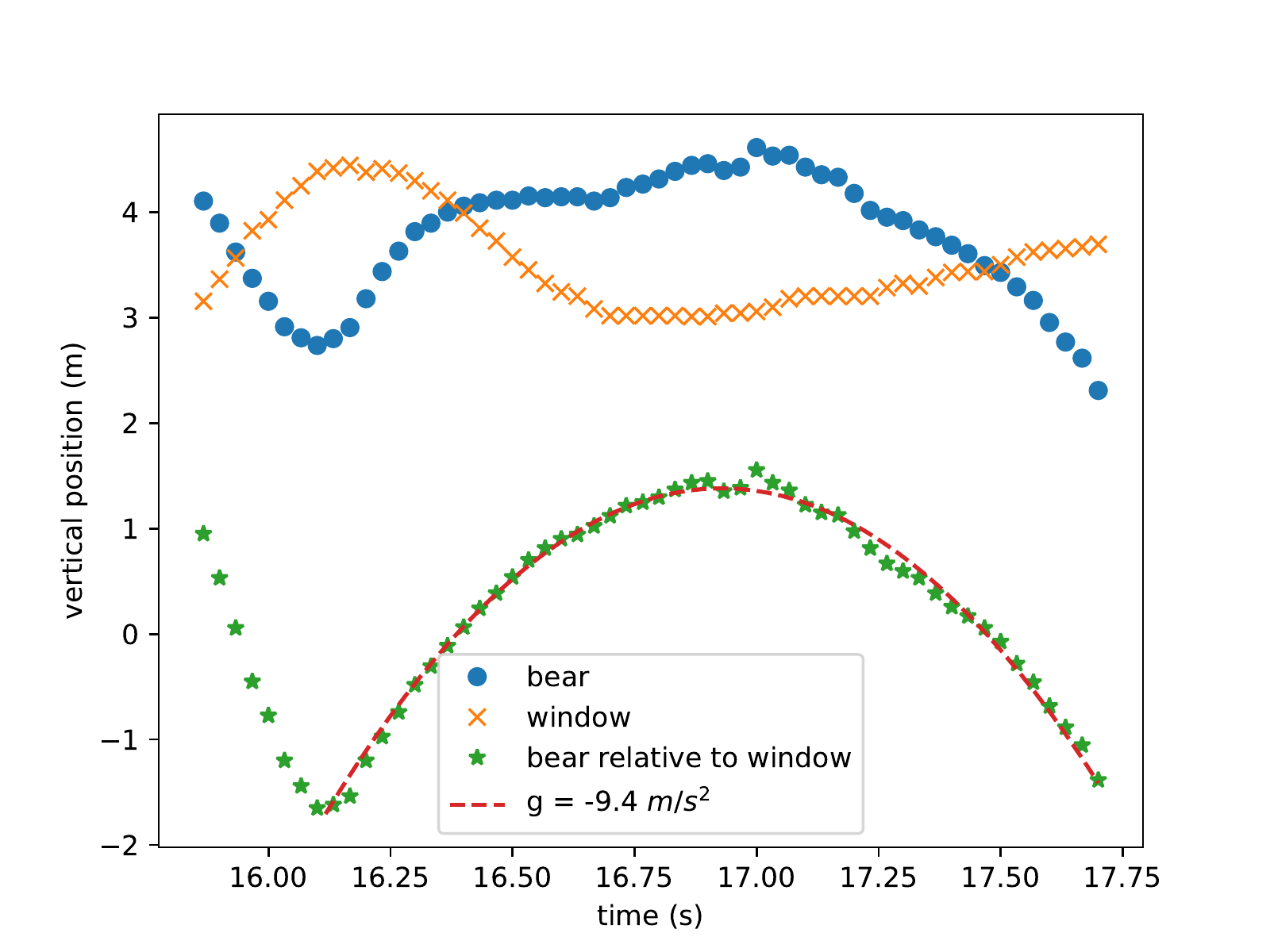}
\caption{
A quadratic fit to the bear's position (relative to the window) shows a $time^2$ polynomial with gravitational acceleration of about $g=9.4\pm 0.1~m/s^2$.  This is obviously not the standard $g$ value from the textbook and suggests both ``clicking error'' in tracking the motion and calibration error in assuming the trampoline is $1$ meter high.  Never-the-less, the corrected motion (unlike that of the bear from the moving camera) is clearly quadratic, which is a demonstration of the utility of the approach used in this paper.
}
\label{bear-quadratic}
\end{figure}

This analysis is shown in figures \ref{bear-dots}, \ref{bear-speed}, and \ref{bear-quadratic}.
 The analysis assumes a trampoline leg height of $1$ meter, which, based on the extracted gravitational acceleration values, is probably inaccurate.
 
 Straight-line fits to the bear's vertical position give speeds of $12.3m/s\approx27mph\approx44kmph$ as the bear hits the trampoline, and $5.9m/s\approx13mph\approx21kmph$ at about the same altitute after bouncing off the trampoline.  Based on the data in Figure 1b of \cite{AccidentRisk} this corresponds to a reduction of pedestrian fatality risk from $5\%$ to less than $1\%$.  Extrapolating from vehicle fatality data, one could arge that in addition to being good comedy, using a trampoline in this case is humane wildlife management.     
     
\section{Conclusion}
Cellphone video is everywhere in a way that was unimaginable 20 years ago.  Most of the videos that students might take for a kinematics assignment won't be shot from a tripod with a constant depth of field.  Accordingly, talking about how to use topics from Physics to correct ``errors'' in video data collection can be empowering for students.  If there's a fixed reference, a student no longer needs to be told, ``That video is bad, go take it again...''.

\begin{acknowledgments}
The work would not have been possible without Mark Hoyoak's excellent videography.  Thanks also to Eugenia Etkina for introducing me to video analysis many years ago.  Thanks also to Peter Bohacek for his inspiring talks and amazing Direct Motion Video examples.  

\end{acknowledgments}

\end{document}